\documentclass[a4paper,11pt]{article}
\pdfoutput=1 % if your are submitting a pdflatex (i.e. if you have
\usepackage{jcappub} % for details on the use of the package, please
\usepackage[T1]{fontenc} % if needed
\usepackage{footnote}

\usepackage{graphicx}
\usepackage{epsfig}

\usepackage{graphicx,ulem}
\usepackage{longtable}
\usepackage{float}
\usepackage{dcolumn}
\usepackage{graphics,epsfig}
\usepackage{amsmath,amssymb,latexsym,mathrsfs}
\usepackage{bm}
\usepackage{color}

\def\he4{$^4$He}
\def\h2{$^2$H}

\newcommand{\angstrom}{\textup{\AA}}

\begin{document}


\title{Cosmological constraints on the neutron lifetime}

\author{L. Salvati$^1$, L. Pagano$^1$, R. Consiglio$^2$, A. Melchiorri$^1$}
\affiliation{$^1$ Physics Department and INFN, Universit\`a di Roma 
	``La Sapienza'', Ple.\ Aldo Moro 2, 00185, Rome, Italy}
\affiliation{$^2$ Physics Department, Universit\`a di Napoli 
	``Federico II'' and INFN, Sezione di Napoli, Complesso Univ. Monte S. Angelo, Via Cintia,
I-80126 Napoli, Italy}
\emailAdd{laura.salvati@roma1.infn.it}
\emailAdd{luca.pagano@roma1.infn.it}
\emailAdd{rconsiglio@na.infn.it}
\emailAdd{alessandro.melchiorri@roma1.infn.it}

\date{\today}

\abstract{
We derive new constraints on the neutron lifetime based on the recent Planck 2015 observations of temperature 
and polarization anisotropies of the CMB. Under the assumption of standard Big Bang Nucleosynthesis, we show that 
 Planck data constrains the neutron lifetime to $\tau_n=(907 \pm 69) \, [\text{s}]$
at $68 \%$ c.l.. Moreover, by including the direct measurements of primordial Helium abundance of Aver et al. (2015) and Izotov et al. (2014), we show that cosmological data provide the stringent constraints $\tau_n=(875 \pm 19) \, [\text{s}]$ and $\tau_n=(921 \pm 11) \, [\text{s}]$ respectively. The latter appears to be in tension with neutron lifetime value quoted by the Particle Data Group ($\tau_n=(880.3 \pm 1.1) \, [\text{s}]$).
Future CMB surveys as COrE+, in combination with
a weak lensing survey as EUCLID, could constrain the neutron lifetime up to a $\sim 6 \, [\text{s}]$  precision.
}

%\pacs{98.80.Cq, 98.70.Vc, 98.80.Es}
\maketitle

\section{Introduction}\label{intro}

The latest measurements of the Cosmic Microwave Background 
(CMB) anisotropies from the Planck satellite have presented a wonderful confirmation
of the standard $\Lambda$CDM cosmological model \cite{Planck2015-1,Planck2015-par}. 
The CMB is therefore providing a new ``laboratory" where several aspects
of fundamental physics can be tested and constrained.
For example, the new Planck measurements have opened the interesting possibility of 
constraining the rate of the  $d(p,\gamma)^3\text{He}$ reaction when combined with
primordial deuterium measurements and assuming primordial standard
Big Bang Nucleosynthesis (BBN) (\cite{Planck2015-par, DiValentino}). 
These cosmological constraints on nuclear reaction rates 
are comparable, in precision, with current direct experimental limits.
Furthermore, again thanks to the new precise data from Planck,
the best constraints to date on atomic transition rates at last scattering have been obtained.
Most notably, the amplitude of the $2s \rightarrow 1$ two-photon rate
that affects recombination is now determined by Planck data with
a $\sim 10 \%$ precision against the current $\sim 43 \%$ experimental
error (see \cite{Planck2015-par} and references therein).

In this paper we work along this line of investigation, showing that 
current cosmological measurements can provide interesting constraints on 
a key quantity in nuclear physics, i.e. the neutron lifetime.

As it is well known, CMB physics is  highly sensitive to the baryon density $\omega_b$ 
and, to a lesser extent, to the primordial Helium fraction $Y_{\text{p}}$. 
On the other hand, standard BBN can provide constraints on the primordial Helium fraction, once the baryon
density and the neutron lifetime $\tau_n$ are fixed. At the same time, one can change perspective and use BBN
to constrain the neutron lifetime, using the constraints on the primordial Helium fraction and
the baryon density coming from CMB data. Clearly, this CMB+BBN constraint on $\tau_n$ can be further improved by  considering the current estimates of $Y_{\text{p}}$ coming from astrophysical observations. 

The obtained constraints are, obviously, model-dependent. They relies not only on the assumption of standard BBN (see, e.g. \cite{Iocco}), but also on the several assumptions made under $\Lambda$CDM. Among them, probably the
most important one is the assumption of a perfect knowledge of the amount of energy density in relativistic
particles both at BBN and CMB epochs. This quantity, often parametrized by the number of relativistic
degrees of freedom $N_{\text{eff}}$ (see e.g. \cite{Planck2015-par}), could be different from the standard value 
of $N_{\text{eff}}=3.046$ if sterile neutrinos, axions or exotic light particles are present and 
could clearly affect the cosmological constraint on $\tau_n$.

On the other hand, this kind of analysis is rather timely since there is still no complete agreement on the experimental value of  $\tau_n$  (for recent and comprehensive reviews see \cite{Pignol,Wietfeldt1,Wietfeldt2}). Indeed, while much technical progress has been made, measured lifetimes can vary by about a percent,
depending on the experimental technique (see e.g. \cite{paul,Pignol}). Two main experimental approaches are available to measure the neutron lifetime: the ``bottle method'' and the ``beam method''. 
In the ``bottle method'' a bottle is filled with Ultra Cold Neutrons (UCN), which are then emptied into a detector to count the number of remaining neutrons. Combining the five most recent UCN measurements \cite{Arzumanov,Steyerl,Pichlmaier,Serebrov,Mampe} one obtains the very precise constraint of $\tau_n^{\text{bottle}}=(879.6 \pm 0.8) \, [\text{s}]$ (see \cite{Pignol}).
On the other hand, in the ``beam method'' detectors record the $\beta$-decay products in a part of the neutron beam. Combining the two most recent measurements \cite{Yue,Byrne} based on the
``beam method'', the following constraint is derived: $\tau_n^{\text{beam}}=(888.0 \pm 2.1) \, [\text{s}]$ (again, see \cite{Pignol}). As we can see, these two averages are in disagreement by $\sim 8.5$ s, i.e. by about $\sim 3.7$ standard deviations.  Moreover, in over twenty years of experimental measurements of $\tau_n$ the average value has fluctuated over several standard deviations (for example, see discussion in \cite{PDG}). The current experimental value quoted by the Particle Data Group (PDG hereafter) of $\tau_n = (880.3 \pm 1.1) \, [\text{s}]$ (see \cite{PDG})  is an average over the best  seven measurements, inflating the error by a scale factor of 1.9.

It is therefore extremely interesting to investigate if current and future cosmological
measurements could constraint the neutron lifetime to a precision that could be useful
to shed light on present experimental discrepancies.
At the same time, this analysis can show to what extent a precise experimental
determination on $\tau_n$ could impact current and future cosmological constraints.

Our work is organized as follows: in the next section we present our analysis method and we 
report the constraints on $\tau_n$ obtained from the Planck 2015 CMB data release. 
We also explore how current measurements of primordial Helium abundances can improve the bounds.
In Section III we forecast the precision achievable by future CMB satellite missions while 
in Section IV we derive our conclusions.

\section{Current constraints on the neutron lifetime}\label{sec:CMB}

\subsection{Method}

We consider the latest CMB anisotropy data coming from the Planck 2015 release (including polarization)
and combinations of this dataset with CMB lensing (\cite{Ade:2015zua}) and Baryonic Acoustic Oscillation (BAO) surveys
(\cite{Anderson}). Even if systematics may be clearly present (as indicated by the tension in the
quoted values, see discussion below) we also consider the inclusion of recent astrophysical measurements of $Y_p$.

\begin{figure}[H]
\centering
\includegraphics[scale=0.6]{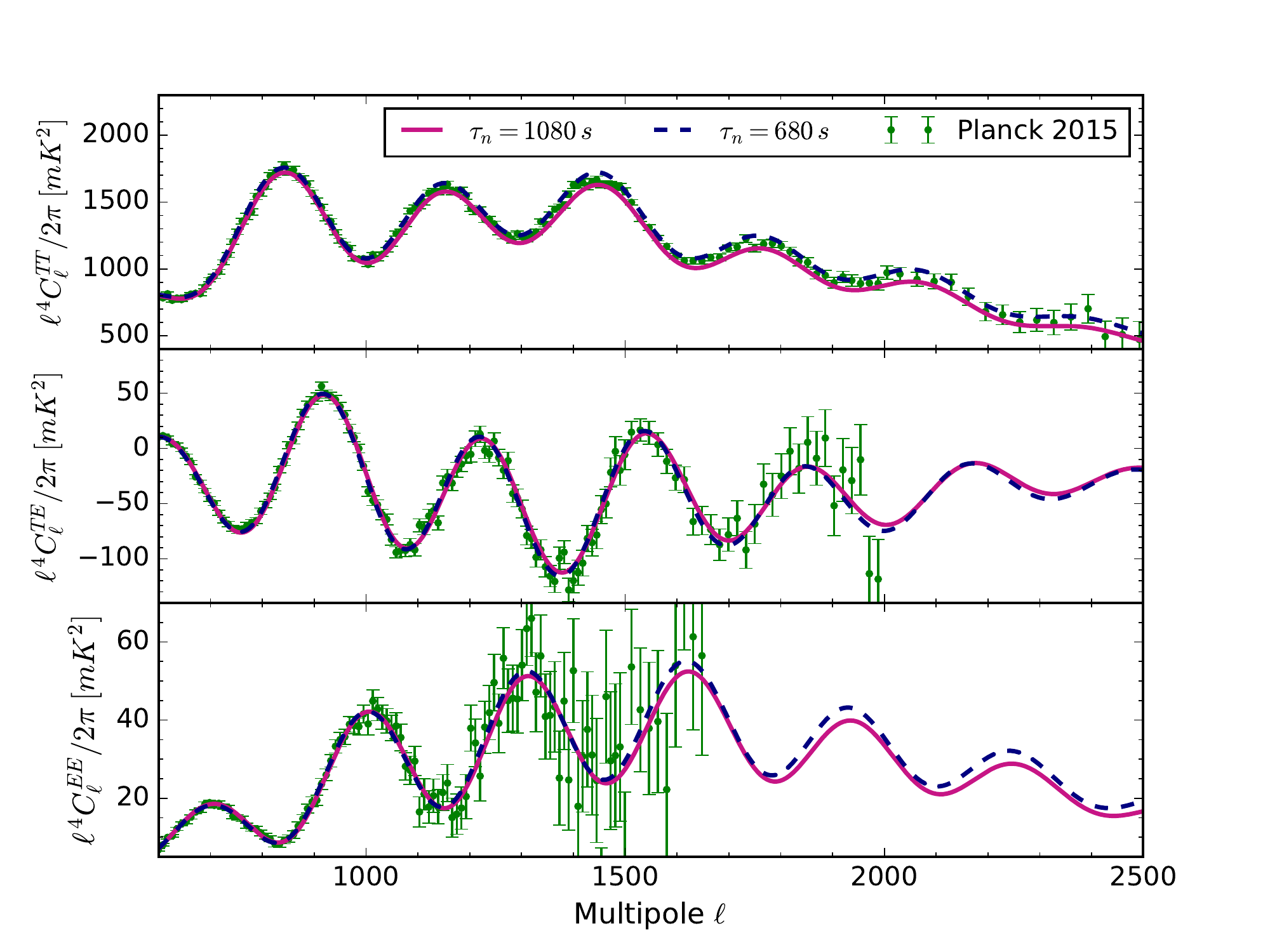}
\caption{\footnotesize{Dependence of CMB power spectra on different values of $\tau_n$, over plotted with the Planck 2015 spectra. }}
\label{fig:spectra}
\end{figure}

The analysis is based on the publicly available Monte Carlo Markov Chain package \texttt{cosmomc} \cite{Lewis:2002ah} which relies on a convergence diagnostic based on the Gelman and Rubin statistic. We use the July 2015 version which includes the support for the Planck Likelihood Code 2.0 \cite{PlanckLikelihood} (see \url{http://cosmologist.info/cosmomc/}) and implements an efficient sampling of the space using the fast/slow parameters decorrelation \cite{Lewis:2013hha}.
We consider the $\Lambda$CDM model, we vary the six standard parameters: the baryon and cold dark matter densities $\omega_b$ and $\omega_c$, the ratio of the sound horizon to the angular diameter distance at decoupling $\theta$, the reionization optical depth $\tau$, the scalar spectral index $n_S$, and the overall normalization of the spectrum $A_S$ at $k=0.05\, \text{Mpc}^{-1}$. We assume adiabatic initial conditions and we impose spatial flatness.
In our baseline analysis we assume standard radiation content (i.e. $N_{\text{eff}}=3.046$), but we also consider
the possibility of relaxing this assumption. As extra parameter, we consider the neutron lifetime $\tau_n$. By itself, CMB is insensitive
to $\tau_n$ but it depends on the primordial Helium abundance $Y_{\text{p}}$ (see e.g. \cite{maria}).
For each step in the chain, we compute a primordial abundance $Y_{\text{p}}^{\text{BBN}}(\omega_b,\tau_n)$ 
obtained using a fitting formula based on the \texttt{PArthENoPE} BBN code~\cite{parthenope}.
Fixing the Helium abundance to $Y_{\text{p}}^{\text{BBN}}$ therefore makes the 
CMB angular spectra sensitive to a variation in $\tau_n$ as we can see in Figure \ref{fig:spectra}.

We present constraints on $\tau_n$ from cosmological data only and also in combination with $^4\text{He}$ direct astrophysical observations. In the last decades many authors \cite{Olive:2004kq, Peimbert:2007vm, Izotov:2007ed, Izotov:2010ca, Izotov:2013waa, Aver:2013wba, Izotov:2014fga, Aver:2015} have attempted to extrapolate primordial abundance of $^4\text{He}$ mainly from the observations of He I emission lines from low-metallicity HII regions. For our analysis, we select results from Izotov et al. 2013 \cite{Izotov:2013waa}, Aver et al. 2013 \cite{Aver:2013wba}, Izotov et al. 2014 \cite{Izotov:2014fga} and Aver et al. 2015 \cite{Aver:2015}. The first two papers represent the most recent results on $Y_{\text{p}}$ from the analysis of optical emission lines, updated with recent set of He I emissivities tabulated by \cite{Porter:2012yb} and larger sample of observed spectra from HII regions. Results from Izotov et al. 2014 and Aver et al. 2015 are the first results with a pioneeristic use of the NIR He I line at $\lambda = 10830 \angstrom$. For detailed description of the extrapolation of $Y_{\text{p}}$ and how they improve previous works see \cite{Izotov:2013waa, Aver:2013wba, Izotov:2014fga, Aver:2015}.

Of course, a key point in our analysis is the precision to which primordial abundance of $Y_{\text{p}}$ is obtained assuming SBBN and given the values of $\omega_b$, $N_{\text{eff}}$ and $\tau_n$.  In this work we rely on the \texttt{PArthENoPE} BBN code, that has been also used by Planck collaboration. However, we need to stress that other BBN codes are available and they may give slightly different values of primordial abundances. A recent comparison between available BBN codes has been made in \cite{Cyburt:2015mya}. Comparing the value of $Y_{\text{p}}$ quoted in Tab. II of \cite{Cyburt:2015mya} with the version of \texttt{PArthENoPE} used in this analysis, we found a $\Delta Y_{\text{p}}=0.0003$, that corresponds to an
accuracy of about $\Delta \tau_n \sim 1.5 \, [\text{s}]$. As we will show later, this level of accuracy is acceptable when considering
current constraints but it will clearly be an issue with future cosmological data.
A second important point is the precision of the BBN fitting formula assumed in our analysis. In principle, for each values of $\omega_b$, $N_{\text{eff}}$ and $\tau_n$ one should run a BBN code to compute the primordial abundances of light elements. However, a single run at fixed parameters is computationally rather expensive and it would be unrealistic to integrate it in the MCMC parameter exploration. For this reason, in BBN calculations, polynomial fitting formulas are used (for details see \cite{Iocco, Planck2015-par}).   Since the formula used in this paper is valid for values of $\tau_n$ close to the PDG one \cite{PDG}, we checked its precision for values within $\Delta \tau_n=100 [\text{s}]$. Given a perfect knowledge of $\omega_b$ and $Y_{\text{p}}$, the expected precision on $\tau_n$ is $0.1 [\text{s}]$ for $\Delta \tau_n=20 [\text{s}]$, $1.3 [\text{s}]$ for $\Delta \tau_n=60 [\text{s}]$ and $3.0 [\text{s}]$ for $\Delta \tau_n=100 [\text{s}]$. We can then infer that these errors are negligible with respect to the statistical ones.\\

\subsection{Results from cosmological and astrophysical data assuming $N_{\text{eff}}=3.046$.}

We first report our constraints assuming a standard energy content in relativistic particles, i.e. $N_{\text{eff}}=3.046$. The results of our analysis are reported in Table \ref{table:CMB-taun} where we report not only the constraints on $\tau_n$ but
also, for completeness, the constraints on $Y_{\text{p}}^{\text{BBN}}$. We indeed point out that these constraints are obtained assuming a complete ignorance on the experimental value of the neutron lifetime.
As we can see from Table \ref{table:CMB-taun}, combining the low$-\ell$ temperature and polarization likelihood \cite{PlanckLikelihood} with the small-scale temperature only likelihood, we obtain the constraint of 
$\tau_n=(918 \pm 105) \, [\text{s}]$. Including the small-scale polarization data from Planck provides a significant
improvement, with $\tau_n = (907 \pm 69) \, [\text{s}]$. Further including to the CMB data  BAO observations \cite{Anderson} we obtain $\tau_n = (915 \pm 63) \, [\text{s}]$, while including also the lensing likelihood \cite{Ade:2015zua} we find $\tau_n = (894 \pm 63) \, [\text{s}]$.

\begin{table}[!h]
\begin{center}
\scalebox{0.8}{
\begin{tabular}{|c||c|c|}
\hline
\rule[-2mm]{0mm}{6mm}
Dataset &$ Y_{\text{p}}^{\text{BBN}}$ &$ \bf \tau _n \, [\text{s}]$ \\
\hline
\hline
\rule[-2mm]{0mm}{6mm}
 Planck TT  & $ 0.254 \pm 0.021$   &  $\bf 918 \pm 105 $  \\
\hline
\rule[-2mm]{0mm}{6mm}
 Planck TT,TE,EE & $ 0.252 \pm 0.014$  & $\bf 907 \pm 69$  \\
\hline
\rule[-2mm]{0mm}{6mm} 
Planck TT,TE,EE + BAO   &  $0.254 \pm 0.013$  & $\bf 915 \pm 63$  \\
\hline
\rule[-2mm]{0mm}{6mm}
Planck TT,TE,EE + BAO + lensing & $0.249 \pm 0.013$ & $\bf 894 \pm 63$  \\
\hline
\end{tabular}}
\caption{\footnotesize{Constraints on $ Y_{\text{p}}^{\text{BBN}}$ and $\tau_n$ at $68 \%$ c.l. based on the Planck 2015 release  (see text).}}
\label{table:CMB-taun}
\end{center}
\end{table}

The constraints on $\tau_n$ coming from current cosmological data are in agreement with the PDG value but  between one and two orders of magnitude weaker than those obtained by laboratory experiments. While future CMB measurements, as we discuss in the next section, could significantly improve these constraints,  is also important to stress that the actual constraints are anyway obtained on very different scales in space and time. Assuming as current value for the neutron lifetime the PDG value of 
$\tau_n^{\text{PDG}}=(880.3 \pm 1.1) \, [\text{s}]$  we see that the Planck data alone constrain any variation in
$\tau_n$ from this quantity to be smaller than  $11 \%$ at $68 \%$ c.l. over a timespan of more than $13.5$ Gyrs.

%\begin{figure}
%\centering
%%\includegraphics[scale=0.210]{figures/test_taun_omegabh2.pdf} %\includegraphics[scale=0.210]{figures/test_taun_ns.pdf}
%\includegraphics[scale=0.245]{figures/2d.png}
%\caption{\footnotesize{Two-dimensional marginalized posterior distributions on the $\tau_n-\Omega_bh^2$ and $\tau_n-n_s$ planes. We show Planck TT, TE, EE + lowP (Planck) alone and in combination with BAO and lensing.}}
%\label{fig:2d}
%\end{figure}

We now provide cosmological constraints on the neutron lifetime by combining 
the latest CMB observations from Planck (including polarization) with four current astrophysical estimates of primordial $Y_{\text{p}}$ described previously.
We combine those values with the Planck full dataset treating them as a gaussian prior on the input Helium abundance and running the MCMC code as described above. We refer to the experimental direct measurement of primordial Helium as $Y_{\text{p}}^{\text{data}}$. 

In Table \ref{table:Yp} we show the results of combining the CMB data with the astrophysical determinations of $Y_{\text{p}}$. 
We report the experimental value of Helium considered ($Y_{\text{p}}^{\text{data}}$), the Helium mass fraction recovered from the MCMC chain ($Y_{\text{p}}^{\text{BBN}}$) and the values for the neutron lifetime ($\tau _n$) for all the datasets analyzed. 
As expected, the constraints on $Y_{\text{p}}^{\text{BBN}}$ are now completely determined by the measured value of
$Y_{\text{p}}^{\text{data}}$.\\
The combined constraining power of the CMB data, sensitive to the baryon density, and the Helium astrophysical measurements allows to tight the bounds on $\tau_n$ by nearly one order of magnitude with respect to the CMB only constraint.

\begin{table}[!h]
\begin{center}
\scalebox{0.8}{
\begin{tabular}{|c||c|c|c|}
\hline
\rule[-2mm]{0mm}{6mm}
Dataset & $Y_{\text{p}}^{\text{data}}$ &$ Y_{\text{p}}^{\text{BBN}}$ &$ \bf \tau _n \, [\text{s}]$ \\
\hline
\hline
\rule[-2mm]{0mm}{6mm} 
Planck TT,TE,EE + Izotov et al. (2013)  & $ 0.254 \pm 0.003$ & $ 0.2539 \pm 0.0029$  & $\bf  916 \pm 15$  \\
\hline
\rule[-2mm]{0mm}{6mm}
Planck TT,TE,EE + Aver et al. (2013) & $ 0.2465 \pm 0.0097  $ &  $ 0.2484  \pm 0.0079 $ &  $ \bf 888  \pm 39  $ \\
\hline
\rule[-2mm]{0mm}{6mm}
Planck TT,TE,EE + Izotov et al. (2014)  & $ 0.2551 \pm 0.0022 $ &  $ 0.2550 \pm 0.0022$ &  $\bf 921 \pm 11 $ \\ 
\hline
\rule[-2mm]{0mm}{6mm}
 Planck TT,TE,EE + Aver et al. (2015) & $ 0.2449 \pm 0.0040  $ &  $  0.2455 \pm 0.0038 $ &  $ \bf 875  \pm 19  $ \\
\hline
\end{tabular}}
\caption{\footnotesize{Experimental priors $ Y_{\text{p}}^{\text{data}}$ from different datasets of direct astrophysical observations. For each datasets we report the obtained value of Helium mass fraction $Y_{\text{p}}^{\text{BBN}} $ and the neutron decay time $\tau_n$.}}
\label{table:Yp}
\end{center}
\end{table}

\begin{figure}[H]
\centering
\includegraphics[scale=0.65]{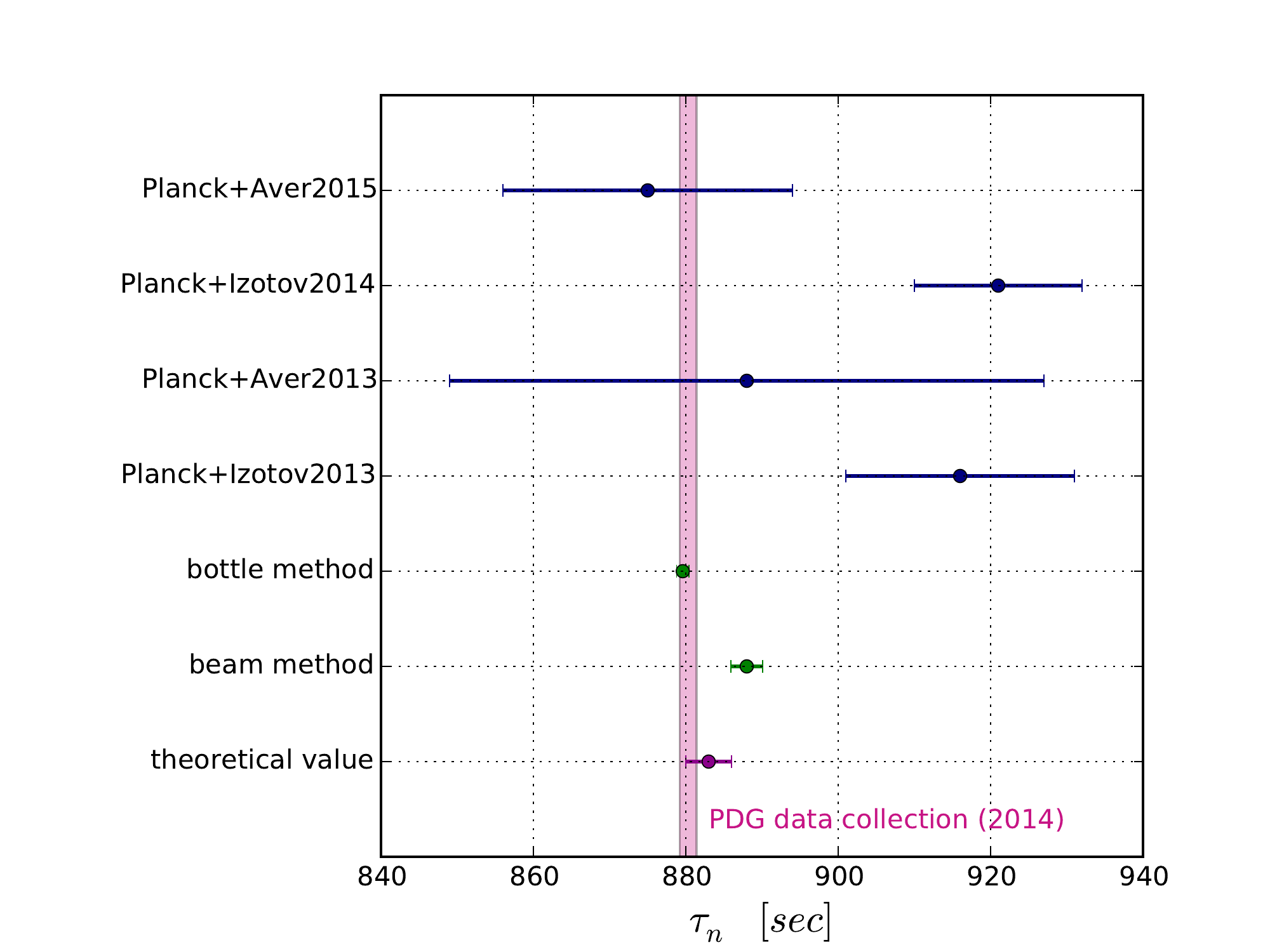}
\caption{\footnotesize{Values of $\tau_n$, all reported with $1 \, \sigma$ error bars. In magenta we report the theoretical value. In green we report the combined values for the ``bottle method'' and ``beam method'' \cite{Pignol}, the vertical violet stripe is the combined value \cite{PDG}. In blue we report all values found including in the analysis the astrophysical measurements.}}
\label{fig:pdg_yp}
\end{figure}

In Figure \ref{fig:pdg_yp} we show all the values for the neutron lifetime evaluated considering individually the astrophysical measurements.
We also show in the same figure the combined values for the ``bottle method'' and ``beam method'' \cite{Pignol}, which are all used in the value quoted by the PDG \cite{PDG}, i.e. $\tau_n = (880.3 \pm 1.1) \, [\text{s}]$, reported in the plot as well. 

Focusing on most recent astrophysical observations, we note that with Izotov et al. (2014) we obtain a value of $\tau_n$ in tension with the PDG one (about $3.7 \,\sigma$), while with the very recent Aver et al. (2015) measure we get $ \tau_n= (875 \pm 19)$ [s], in perfect agreement with the PDG but still not sensitive enough to discriminate between ``bottle method'' and ``beam method''.
The Izotov et al. (2014) and Aver et al. (2015) datasets provide values of $\tau_n$, when combined with CMB data, that
are in tension at the level of $2.1 \, \sigma$. This is expected since it reflects the tension between the two primordial abundance extrapolations.

\subsection{Results from cosmological and astrophysical data obtained varying $N_{\text{eff}}$.}

As mentioned above, we now check how our constraints are dependent on the chosen assumptions for the dark radiation content. Therefore we include $N_{\text{eff}}$ as extra parameter, performing the analysis both with cosmological data alone and in combination with primordial Helium abundance observations. 
Moreover it would be interesting to investigate if the inclusion of $N_{\text{eff}}$ could alleviate the tension on the value of $\tau_n$ between Izotov et al. (2014) and Aver et al. (2015).
%For this refined analysis we consider only the two most recent astrophysical observations of Izotov et al. (2014) and Aver et al. (2015) since is interesting to investigate if the inclusion
%of $N_{\text{eff}}$ could alleviate the tensions on the value of $\tau_n$ reported above.

\begin{figure}[H]
\centering
\includegraphics[scale=0.65]{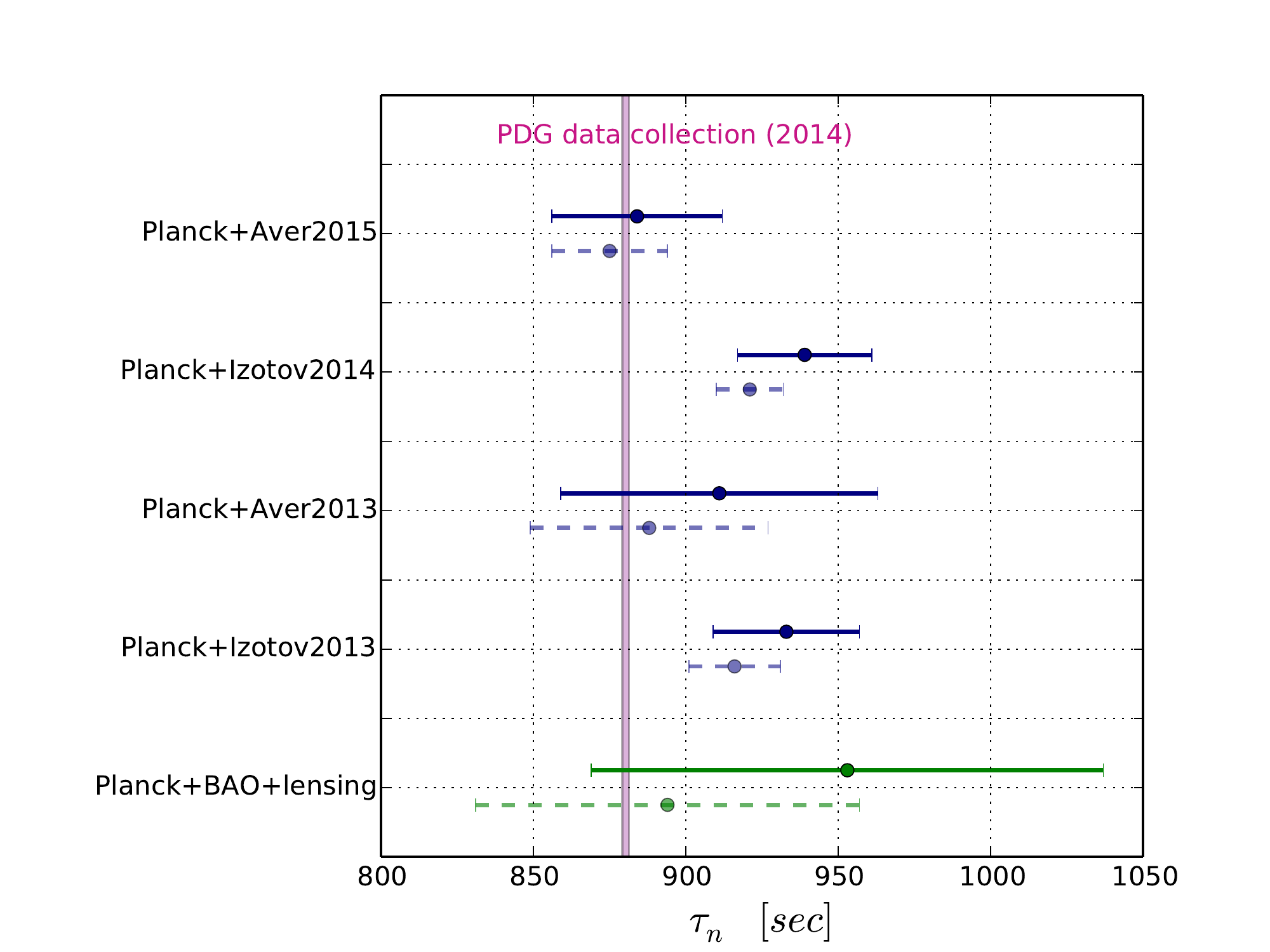}
\caption{\footnotesize{Values of $\tau_n$, all reported with $1 \, \sigma$ error bars. The vertical violet stripe is the combined value \cite{PDG}. In green we report values obtained using cosmological data and in blue results from the combination of Planck TT,TE,EE with astrophysical observations of primordial Helium abundance. For comparison, we report results obtained fixing $N_{\text{eff}}=3.046$ (dotted lines) and relaxing this assumption (solid lines).}}
\label{fig:taun_neff}
\end{figure}

The results of this analysis are shown in Table \ref{table:Neff} and in Figure \ref{fig:taun_neff}. As expected we see an overall  relaxation of the constraints on $\tau_n$ (a factor of about $1.3$ for the cosmological constraints and about $1.5 \div 2$ for the
combined CMB and astrophysical measurements analysis), due to the well known degeneracy between $N_{\text{eff}}$ and $ Y_{\text{p}}$. Given this degeneracy and considering that cosmological data now prefer values of $N_{\text{eff}}$ {\it lower} than the standard one, we obtain higher primordial Helium abundances and, as direct consequence, higher values of $\tau_n$. Despite this overall behaviour, these new results are in agreement within $1 \, \sigma$ with those discussed above (obtained fixing $N_{\text{eff}}=3.046$), as shown in Figure \ref{fig:taun_neff}. In addition we always obtain values of $N_{\text{eff}}$ in agreement with the standard one within $1 \, \sigma$.
As we can see, the tension between Aver et al. (2015) and Izotov et al. (2014) persists, albeit now at the level of
$1.5 \, \sigma$. Of course, part of the previous $2 \, \sigma$ discrepancy has been absorbed by the $N_{\text{eff}}$ parameter, resulting in a $\sim 0.5 \, \sigma$ shift between Izotov et al. (2014) and Aver et al. (2015) on $N_{\text{eff}}$ (see Table \ref{table:Neff}).

\begin{table}[!h]
\begin{center}
\scalebox{0.8}{
\begin{tabular}{|c||c|c|c|}
\hline
\rule[-2mm]{0mm}{6mm}
Dataset & $ Y_{\text{p}}^{\text{BBN}}$ & $N_{\text{eff}}$ &$ \bf \tau _n \, [\text{s}]$ \\
\hline
\hline
\rule[-2mm]{0mm}{6mm}
 Planck TT,TE,EE & $0.263 \pm 0.018$ & $2.76 \pm 0.30$ & $\bf  986 \pm 109$ \\
\hline
\rule[-2mm]{0mm}{6mm}
Planck TT,TE,EE + BAO + lensing & $0.258 \pm 0.014$ &  $2.83 \pm 0.25$ & $\bf  953 \pm 84$   \\
\hline
\hline
\rule[-2mm]{0mm}{6mm}
Planck TT,TE,EE + Izotov et al. (2013)  & $  0.2542 \pm 0.0029  $ &  $ 2.84 \pm 0.23$ &  $\bf 933 \pm  24$ \\ 
\hline
\rule[-2mm]{0mm}{6mm}
Planck TT,TE,EE + Aver et al. (2013)  & $  0.2505 \pm  0.0085$ &  $  2.90 \pm 0.25 $ &  $\bf 911 \pm 52  $ \\ 
\hline
\rule[-2mm]{0mm}{6mm}
Planck TT,TE,EE + Izotov et al. (2014)  & $ 0.2552 \pm 0.0022 $ &  $ 2.83 \pm 0.23$ &  $\bf 939 \pm 22 $ \\ 
\hline
\rule[-2mm]{0mm}{6mm}
 Planck TT,TE,EE + Aver et al. (2015) & $ 0.2458 \pm 0.0040  $ &  $  2.95 \pm 0.24 $ &  $ \bf 884  \pm 28  $ \\
\hline
\end{tabular}}
\caption{\footnotesize{Constraints at $68 \%$ c.l. on neutron lifetime obtained varying $N_{\text{eff}}$ in addition to the other cosmological parameters. In the last two rows we combine astrophyisical observations with Planck TT,TE,EE.}}
\label{table:Neff}
\end{center}
\end{table}

\subsection{Theoretical estimate}
For reference, we also show in Figure \ref{fig:pdg_yp} a theoretical estimate for $\tau_n
= (883 \pm 3)$ [s]. This value is the result including QED {\it outer}
corrections (treating nucleons as a whole), and {\it inner} corrections
depending on nucleon structure, the latter promoted at leading log level
as in \cite{marciano}. Perturbative QCD corrections, weak magnetism and 
finite nucleon mass effects, related to proton recoil, as well as the Fermi 
function describing the electron rescattering in the proton Coulomb field, 
are also included. For more details on these points
see e.g. \cite{wilkinson,sirlin,marciano,esposito}. The value of parameters
entering this expression, such as vector and axial coupling $C_V$ and
$C_A$, fine structure constant, particle masses, strong coupling constant
etc. has been updated to the most recent results reported in \cite{PDG}.
Finally, the low energy
cutoff applied to the short-distance part of the $\gamma-W$  box diagram
is chosen in the range $M_A=1-1.5$ GeV. The theoretical error is obtained
by propagating the uncertainties on all parameters to the value of
$\tau_n$, and is largely dominated by the present error on $C_A/C_V$.
Of course, the theoretical estimate is still affected by smaller effects,
such as for example, higher order terms in $\alpha$, sub-leading
corrections, residual average proton polarization due to parity
non-conservation and so on \cite{wilkinson}. All these further corrections are
expected to be quite sub--dominant \cite{Marciano:2005ec}.\\

\section{Forecasts on future data}

Considering that current cosmological and astrophysical observations are not competitive with particle physics experiments yet and they also could be affected by unknown systematics, we now perform forecasts on forthcoming observations.

First af all, it is interesting to  evaluate the precision on Helium abundance from future astrophysical observations, combined with Planck, needed to reach the PDG accuracy. We find that to achieve 1 second precision on $\tau_n$ we need to measure $Y_{\text{p}}$ with an error of $0.0002$, roughly an order of magnitude better than the current measurements. This would need a better control of the systematic errors on direct astrophysical measurements.
Moreover, as discussed in the previous section, this would also need a significant, but in our opinion
achievable, improvement in the numerical accuracy of current BBN codes, see e.g. a recent paper on possible corrections due to a better treatment of neutrino energy transport \cite{Grohs:2015tfy}.

\begin{table}[!h]%\footnotesize
\begin{center}
\scalebox{0.85}{
\begin{tabular}{|r|c|c|c|c|c|}
\hline
\rule[3mm]{0mm}{0mm}
Experiment & $f_{\text{sky}}$ & Channel & FWHM & T $\mu K \cdot$ arcmin & Q/U $\mu K \cdot$ arcmin \\
\hline
\hline
\rule[3mm]{0mm}{0mm}
COrE & 0.80  & 105 & 10' & 2.68 & 4.63\\
 & & 135 & 7.8' & 2.63 & 4.55 \\
 & & 165 & 6.4'& 2.67 & 4.61\\
 & & 195 & 5.4'& 2.63 & 4.54\\
 & & 225 & 4.7'& 2.64 & 4.57\\
\hline
\rule[3mm]{0mm}{0mm}
AdvACT &  0.50 & 90 & 2.2' & 7.8 & 10.9   \\
&  &  150  & 1.3'   & 6.9  &  9.7 \\
 & &  230  & 0.9'  & 25  & 35  \\
 \hline
\rule[3mm]{0mm}{0mm}
SPT-3G &  0.06 & 95 &1.6 ' & 4.2 & 5.9\\
 &   & 150 & 1.0' & 2.5 & 3.5 \\
 &   & 220 & 0.68' & 4.2 & 5.9\\
 \hline
\rule[3mm]{0mm}{0mm}
CMB-S4 & 0.50 & 150  & 1.3' & 1& 1.4 \\
\hline
\rule[3mm]{0mm}{0mm}
CVL &  1.00 & 150 & 5' & 0 & 0\\
\hline
\end{tabular}}
\caption{\footnotesize{Experimental specifications for COrE \cite{2011arXiv1102.2181T}, AdvACT \cite{Calabrese:2014gwa}, CMB-S4 \cite{Mueller:2014dba}, SPT-3G \cite{SPT-3G} and CVL. We report the used sky fraction $f_{\text{sky}}$, the used channels, the beam FWHM in arc-minutes and the sensitivity per pixel in temperature and polarization. }}
\label{tab:exp}
\end{center}
\end{table}

It is also interesting to evaluate the future constraints
on $\tau_n$ achievable from an analysis based just on cosmological data as CMB or galaxy clustering.
CMB sensitivity on $\tau_n$ is encoded in the small-scale region of the power spectrum, therefore we can expect tighter constraints from next generation CMB missions, planned to measure such $\ell$ range. To forecast the sensitivity on $\tau_n$  that future CMB experiments can achieve, we simulate different synthetic datasets. We consider the following future experiments: COrE \cite{2011arXiv1102.2181T}, AdvACT \cite{Calabrese:2014gwa}, Stage IV CMB experiment (CMB-S4 hereafter) \cite{Mueller:2014dba} and SPT-3G \cite{SPT-3G}. All the experimental specifications are reported in Tab. \ref{tab:exp}. Concerning the COrE experiment, we have chosen the 5 channels in the range 100-220 GHz. We also consider an ideal \textit{Cosmic Variance Limited} (CVL hereafter) case, to evaluate the most accurate precision that can be reached with CMB experiments, assuming no systematic effects and variance dominated by the signal in the considered range of multipoles. In order to extract the cosmological parameter constraints we generate simulated datasets following the, now common, approach described in \cite{Lewis:2005tp}. We assume, as fiducial model, the Planck 2015 best fit. Moreover we assume that uncertainties due to foreground removal are smaller than statistical errors, that beam uncertainties are negligible and white noise. As Likelihood function we use the Inverse Wishart distribution,  reducing the degrees of freedom at each $\ell$ by $f_{sky}$ and ignoring the correlations. Results are reported in Table \ref{table:CMB-taun-fore}.

\begin{table}[!h]
\begin{center}
\scalebox{0.8}{
\begin{tabular}{|c||c|c|}
\hline
\rule[-2mm]{0mm}{6mm}
Dataset &$ Y_{\text{p}}^{\text{BBN}}$ &$ \bf \tau _n \, [\text{s}]$ \\
\hline
\hline
\rule[-2mm]{0mm}{6mm}
Planck TT,TE,EE + AdvACT & $0.2464 \pm 0.0065$ & $\bf 879 \pm 32$ \\
\hline
\rule[-2mm]{0mm}{6mm}
Planck TT,TE,EE + CMB-S4 & $ 0.2475 \pm 0.0037$ & $\bf 884 \pm 18$ \\
\hline
\rule[-2mm]{0mm}{6mm}
Planck TT,TE,EE + SPT-3G & $ 0.2487 \pm 0.0091$ & $\bf 890 \pm 44 $ \\
\hline
\hline
\rule[-2mm]{0mm}{6mm}
COrE & $0.2467 \pm 0.0023$ & $\bf 880 \pm 11$ \\
\hline
\rule[-2mm]{0mm}{6mm}
CVL & $ 0.2467 \pm 0.0011$ & $\bf 880.7 \pm 5.5$ \\
\hline
\hline
\rule[-2mm]{0mm}{6mm}
Planck TT,TE,EE + Euclid & $ 0.2521 \pm 0.0069$  & $\bf 907 \pm 34$  \\
\hline
\rule[-2mm]{0mm}{6mm}
COrE + Euclid & $0.2467 \pm 0.0014$ & $\bf 880.3 \pm 6.7 $  \\
\hline
\end{tabular}}
\caption{\footnotesize{Future constraints on $ Y_{\text{p}}^{\text{BBN}}$ and $\tau_n$ at $68 \%$ c.l. achievable from the considered cosmological datasets (see text).}}
\label{table:CMB-taun-fore}
\end{center}
\end{table}

Combining the Planck 2015 likelihood with simulated datasets for the future ground based missions SPT-3G, AdvACT and CMB-S4 the $1 \, \sigma$ error on the measured $\tau_n$ is reduced respectively to   $ 44 \, [\text{s}]$, $32 \, [\text{s}]$ and to $18 \, [\text{s}]$. The next generation satellite mission COrE could lower this constrain to $11 \, [\text{s}]$ while the maximum intrinsic limit achievable with the CMB data alone is about $5.5 \, [\text{s}]$.  Finally, we also perform a forecast for the future Euclid mission \cite{Euclid}, combining it with the latest Planck data and with the future COrE mission. We use the Fisher matrix formalism, following \cite{Amendola} and the experimental specifications reported in \cite{Martinelli}. We sum the Euclid Fisher matrix with the CMB inverse covariance produced by \texttt{cosmomc}, assuming gaussian probability distribution for the cosmological parameters. Despite Euclid being not sensitive to the variation of $\tau_n$, it clearly helps in breaking the degeneracies between the cosmological parameters. Infact we obtain a significative reduction of the error on $Y_{\text{p}}^{\text{BBN}}$ and $\tau_n$ with respect to the Planck only and CorE only cases, as reported in Table \ref{table:CMB-taun}.

\section{Conclusions}\label{Conclusions}

In this paper we have presented new constraints on the neutron lifetime combining
a set of current cosmological data. In particular, we have shown that the recent
Planck 2015 temperature and polarization data release, under the assumption of standard BBN, constrains the neutron lifetime to
$\tau_n = (907 \pm 69) \, [\text{s}]$. While the uncertainty on this value is significantly larger than current experimental estimates, it provides a new and completely independent measurement.

We have combined the Planck constraints with current astrophysical
determinations of $Y_{\text{p}}$. We have found that in this case a 
$\Delta \tau_n \sim 10 \div 20 \, [\text{s}]$ accuracy could be reached. This level of sensitivity is still not able to discriminate between ``beam method'' and ``bottle method''. Anyway, forecasting future constraints achievable by a combination
of Euclid and COrE+ satellite mission, we have shown  that these experiments could
reach,in principle, a $\Delta \tau_n \sim 6\, [\text{s}]$ sensitivity, providing also an improved accuracy in current
BBN codes. Therefore, by combining future and more precise cosmological and astrophysical observations we could a reach 
sensitivity comparable with current experimental uncertainties, with the caveat that all the possible sources of systematics are properly taken into account.

%%%%%%%%%%%%%%%%%%%%%%%%%%%%%%%%%%%%%%%%%%%%%%%
\section*{Acknowledgments}
We would like to thank G. Mangano for useful discussions.
LP and AM acknowledge partial financial support from PRIN 2012 
Fisica Astroparticellare Teorica (Protocollo 
2012CPPYP7\_009, CUP (INFN) I18C13000080001). We also would like to thank the anonymous referee for her/his very useful suggestions during the review process.%
%%%%%%%%%%%%%%%%%%%%%%%%%%%%%%%%%%%%%%%%%%%%%%%
%%%%%%%%%%%%%%%%%%%%%%%%%%%%%%%%%%%%%%%%%%%%%%%

\end{document}